# *Towards contrast- and pathology-agnostic clinical fetal brain MRI segmentation using SynthSeg*

Ziyao Shang[1,2], Misha Kaandorp[1,3], Kelly Payette[1], Marina Fernandez Garcia[1,11], Roxane Licandro[4], Georg Langs[4], Jordina Aviles Verdera[8,9], Jana Hutter[8,9,10], Bjoern Menze[12], Gregor Kasprian[4,5], Meritxell Bach Cuadra[6,7], Andras Jakab[1,3]

[1] Center for MR Research, University Children's Hospital Zurich, Zurich, Switzerland
[2] ETH ZURICH, Department of Computer Science
[3] Faculty of Medicine, University of Zurich, Zurich, Switzerland
[4.] Computerized Imaging Research Group, Department of Biomedical Imaging and Image-guided Therapy, Medical University of Vienna
[5.] Division of Neuroradiology and Musculoskeletal Radiology, Department of Biomedical Imaging and Image-guided Therapy, Medical University of Vienna, Vienna
[6.] CIBM Center for Biomedical Imaging, Lausanne, Switzerland
[7.] Radiology Department, University of Lausanne and Lausanne University Hospital, Lausanne, Switzerland
[8.] Research Department of Early Life Imaging, School of Biomedical Engineering and Imaging Sciences, King's College London, London, UK.
[9.] Biomedical Engineering Department, School of Biomedical Engineering and Imaging Sciences, King's College London, London, UK.
[10.] Smart Imaging Lab, Radiological Institute, University Hospital Erlangen, Erlangen, Germany.
[11.] MRILab, Institute for Molecular Imaging and Instrumentation (i3M), Spanish National Research Council (CSIC), Universitat Politècnica de València (UPV), Valencia, Spain
[12] Department of Quantitative Biomedicine, University of Zurich, Zurich, Switzerland.

[*] **Corresponding Author**
**Name:** Andras Jakab
**Department:** Center for MR Research
**Institute:** University Children's Hospital Zurich
**Address:** Lenggstrasse 30, 8008 Zürich, Switzerland
**Email:** andras.jakab@kispi.uzh.ch

## Abstract

Magnetic resonance imaging (MRI) has played a crucial role in fetal neurodevelopmental research. Structural annotations of MR images are an important step for quantitative analysis of the developing human brain, with Deep Learning providing an automated alternative for this otherwise tedious manual process. However, segmentation performances of Convolutional Neural Networks often suffer from domain shift, where the network fails when applied to subjects that deviate from the distribution with which it is trained on. In this work, we aim to train networks capable of automatically segmenting fetal brain MRIs with a wide range of domain shifts pertaining to differences in subject physiology and acquisition environments, in particular shape-based differences commonly observed in pathological cases. We introduce a novel data-driven train-time sampling strategy that seeks to fully exploit the diversity of a given training dataset to enhance the domain generalizability of the trained networks. We adapted our sampler, together with other existing data augmentation techniques, to the SynthSeg framework, a generator that utilizes domain randomization to generate diverse training data. We ran thorough experimentations and ablation studies on a wide range of training/testing data to test the validity of the approaches. Our networks achieved notable improvements in the segmentation quality on testing subjects with intense anatomical abnormalities ($p < 1e\text{-}4$), though at the cost of a slighter decrease in performance in cases with fewer abnormalities. Our work also lays the foundation for future works on creating and adapting data-driven sampling strategies for other training pipelines.

**Keywords**: MR image segmentation, domain shift, deep neural networks, human brain development, data augmentation.

## 1. Introduction

Magnetic resonance imaging (MRI) has become one of the most commonly used methods for studying the development of the human brain in the clinical and research context in the past decades. Its non-invasive nature and excellent contrast make it a promising modality, even for challenging populations like the human fetus and newborn. Fetal MRI is playing an emerging role in the prenatal counselling process, particularly for characterizing developmental abnormalities affecting the brain (Manganaro et al., 2023; Prayer et al., 2023). To quantify brain development, segmentation - assigning each voxel in fetal MR images to the structure it belongs to - is a crucial step. However, manually acquiring accurate segmentations demands extensive efforts from specialized medical professionals and could be very time-consuming.

Artificial Intelligence (AI)-driven fetal MR image segmentation tools can provide robust and efficient alternatives. Deep learning, in particular convolutional neural networks (CNNs), have revolutionized automated MR image segmentations, resulting in powerful CNN-based segmentation models. For challenging fetal brain MRI datasets, CNNs have become the dominant approach. The Fetal Tissue Annotation (FeTA) MICCAI challenges (Payette et al., 2025, 2023, 2021) provided the first open dataset and benchmarks to test automated fetal brain segmentation algorithms and demonstrated the power of U-Net-based structures (Ronneberger et al., 2015). However, they also revealed the sensitivity of U-Net models to domain shifts, as without proper adaptations, U-nets were not able to produce high quality segmentation when the testing data distributions deviated from the training data (Payette et al., 2025).

Domain shifts in fetal imaging manifest in various ways. In particular, during early brain development, significant structural changes occur in the brain, such as cortical gyrification, cerebellar growth, and lateral ventricle volume and shape change. Fetal brain pathologies (e.g., ventriculomegaly, cerebellar hypoplasia) further alter brain morphology. The fetal brain image for segmentation may or may not include varying amounts of maternal tissues and amniotic fluid. More generally, domain shift may also occur as a result of image-based artifacts such as bias and signal drop-out, as well as differences in MRI contrast and intensity, variations in scanning parameters, magnet strengths, and manufacturers across medical centers. This heterogeneity poses significant challenges to automated fetal MR image segmentation methods.

Utilization of synthetic training data has proven to be an effective strategy for addressing domain shift as it provides high degrees of freedom when generating diverse training data for image segmentation models (Gopinath et al., 2024). Adapting the concept of domain randomization (Tobin et al., 2017), SynthSeg (Billot et al., 2023) provided a pipeline that can generate synthetic image/segmentation pairs with very diverse contrasts and shape deformations for training, often to the extent of appearing very unrealistic. This generation pipeline also has the advantage of only requiring segmentation maps (i.e. training templates) and no actual MR images as input. Training data generated by SynthSeg has been able to train U-Net-based segmentation models with good domain adaptability. Previous efforts to apply SynthSeg to brains in the early developmental stages have primarily focused on structural subdivisions (Shang et al., 2022; Zalevskyi et al., 2024) and simulating motion and MRI artifacts (Valabregue et al., 2023). However, they have not been applied to pathological fetal brains, which very commonly occur in the clinical practice of fetal MRI diagnostics.

The aim of our work is to enhance the generalizability of segmentation networks to shape-based domain shifts in severely pathological cases. We also aim to evaluate their ability to generalize for domain shifts arising from variations in MRI contrast, imaging settings, and post-processing pipelines - challenges commonly encountered in clinical fetal MRI. We introduce a novel, data-driven sampling strategy for training templates based on their shape representations, which prioritizes diversity in anatomical and pathological variability when selecting training templates. This is because fetal MRI is typically limited by a relatively (i.e., compared to ultrasonic images) smaller amount of biometric reference data (Di Mascio et al., 2022), mostly acquired in clinical settings and are not representative of cross-sectional, population-based studies. Thus, our strategy aims to maximize the effectiveness of an often-limited training set. We combine our sampling strategy with binary morphological operations adopted from existing work (Kaandorp et al., 2025), which deform the brain's anatomical structure and tailor them to resemble real-life pathologies, seeking to improve the network's generalizability across a wider variety of brain conditions. Such structure-specific deformations complement the augmentations performed by SynthSeg, which only conducts global (i.e., structural-agnostic) deformations. Our strategy is also combined with background subdivision, where parts not belonging to any brain structure are divided into sub-groups and augmented differently when training data is generated (Zalevskyi et al., 2024).

Those combinations are evaluated by a comprehensive series of ablation studies on a wide range of different training and testing datasets from various hospitals, including pathological subjects with significant structural shape deviations from neurotypical subjects, different image contrasts, MRI sequences, imaging setting (pre- vs. postnatal preterm), image resolutions and super-resolution reconstruction methods. Thus, our work also serves as a validation to those existing methodologies on U-net segmentation models trained by the SynthSeg generator. Our results showcase the good generalizability of SynthSeg and its combination with such methodologies. Our data-driven sampling strategy achieved particularly notable improvements on testing data with large shape variances, though with a trade-off reflected in a slight performance decrease on data with fewer structural irregularities. Last, our data-driven sampler is compared with representative state-of-the-art works related to domain shift and fetal brain MRI segmentations, including FetalSynthSeg (Zalevskyi et al., 2024), the FeTA challenge series (Payette et al., 2025, 2023, 2021; Zalevskyi et al., 2025a), and DRIFTS (Zalevskyi et al., 2025b). The results further illustrate the advantages of data-driven sampling on subjects with significant shape differences in anatomical structures.

In summary, in this work we introduce a data-driven sampling strategy that, to our knowledge, is the first to explicitly balance structural shape variability in brain segmentation training sets, addressing a limitation in current pipelines. We also extend SynthSeg by adding structure-specific morphological operations that generate anatomically targeted deformations, complementing SynthSeg's reliance on global, structure-agnostic transformations. Finally, we evaluate these contributions across a large and diverse collection of clinically used, real-life datasets divided into multiple imaging domains, demonstrating that our methods improve cross-domain generalizability beyond what SynthSeg alone can achieve.

## 2. Methods

### 2.1 Datasets

In this section, we introduce details regarding the curation and partition for datasets used in our experiments. Section 2.1.1 focuses on the collection, processing, and organization of training/testing signal maps. Section 2.1.2 describes how the ground truth structural segmentation maps used in our experiments are generated.

### 2.1.1 MRI datasets and rationale of data pooling for test diversification

An illustration of our datasets and domain divisions can be found in Figure 1, with their metadata information listed in Table 1. Below we briefly describe all datasets used and how they are grouped.

**Training datasets:** We used two openly available fetal MRI datasets for training data: the FeTA 2021 Challenge dataset (FeTA Challenge 2021 (Payette et al., 2023), University Children's Hospital Zürich, Switzerland) and the dHCP dataset (Hughes et al., 2017; Makropoulos et al., 2018) (dHCP fetal release, UK). Only the anatomical segmentation templates (label maps) are required by SynthSeg for training.

**Testing datasets:** We used a broad range of testing datasets from different medical centers and image modalities. The following considerations were made to pool testing data that represents in-domain, similar-domain and out-of-domain data, classified based on the subjective degree of domain shift experienced compared to the training data. An additional *most-pathological* "domain" is also created to test the network's efficacy on subjects with the most severe ventriculomegaly (i.e., shape differences). The main domain-shift factors we consider for our data pooling scheme are acquisition, reconstruction, modality, and structural shape variations (i.e., due to pathologies).

1. **In-domain data** ("FeTA test") was considered as data identical to the training set, meaning T2-weighted, super-resolution reconstructed datasets from the FeTA Challenge test set, acquired on the same MRI scanners as the training data and using the same image processing pipeline (SVRTK (Uus et al., 2020) and MIAL-SRTK (Tourbier et al., 2015)), and a similar proportion of normal and pathological subjects (the test data was not released openly).

2. **Similar-domain data** represented fetal MRI data acquired in the same hospital as the training data (FeTA 2021), however, either undergoing a different image reconstruction pipeline ("ZRH new crtl", using NesVor (Xu et al., 2023)) or being synthetically generated using a SPADE GAN based label to MR image generator network (Garcia et al., 2022) ("ZRH synth T2"), which represents a domain with different image features.

3. **Out-of-domain data** was diversified by incorporating types of datasets commonly used in clinical and research studies involving fetal MRI or postnatal MRI in preterm infants ("dHCP pretT2" and "dHCP pretT1") over recent years. Although fetal MRI primarily relies on T2-weighted imaging, improving segmentation accuracy in T1-weighted images remains a challenge due to domain shift, and matched T1-T2 image pairs are often unavailable in the clinical reality in good image quality that would enable super-resolution reconstruction. To introduce pronounced image contrast shift, we generated synthetic fetal T1 images ("ZRH synth T1") from the T2-weighted images in the FeTA dataset using a Pix2Pix network described previously (Moatti, 2022). To introduce scanner field strength related domain shift, data from a low-field 0.55T scanner was used ("KCL" from King's College London, UK). To introduce domain shifts due to a different MRI protocol and image reconstruction, data from two further hospitals were included: Lausanne University Hospital ("CHUV"), Lausanne, Switzerland and Medical University of Vienna, Austria ("Vienna").

4. **Most-pathological data**: For an evaluation of the most pathological cases with severe ventriculomegaly, a sub-group consisting of 33 subjects from the FeTA 2021 test set, the "ZRH Spina" set, and the "Vienna" dataset were sampled (examples shown in Figure 2) to form this additional domain.

Regarding shape-based domain shifts, we aimed to account for both neurotypical cases and fetuses with ventriculomegaly and structural abnormalities associated with spina bifida, a common clinical indication for fetal MRI, to ensure robust segmentation performance across a diverse range of anatomical variations ("ZRH Spina", University Children's Hospital Zürich). Furthermore, we sought to evaluate whether the segmentation network is able to generalize for shape-based domain shift introduced by birth by using postnatal preterm T1 and T2-weighted data of infants from the dHCP Neonatal Release 2 with comparable gestational age as the older subjects in the FeTA datasets.

Table 1: General information about datasets used for training and testing.

| Domain | Dataset | Data Source* | n | Age at scan (GA***, weeks): mean ± SD (range) | n_normal/ n_pathological | MRI Field strength | MRI contrast | Resolution ($mm^3$) | Reconstruction method |
|---|---|---|---|---|---|---|---|---|---|
| Training | FeTA train | 1 | 80 | 27.0 ± 3.6 (20.0-34.8) | 31 / 49 | 1.5 / 3.0T | T2 | 0.5 | MIAL-SR, IRTK |
| | dHCP fetal train | 2 | 265 | 28.7 ± 3.9 (21.1-38.3) | 265 / 0 | 3.0T | T2 | 0.5 | SVRTK |
| Test: in-domain | FeTA test | 1 | 40 | 27.0 ± 3.4 (21.3-34.6) | 15 / 25 | 1.5 / 3.0T | T2 | 0.5 | MIAL-SR, IRTK |
| Test: similar-domain | ZRH new crtl | 3 | 44 | 27.8 ± 3.8 (20.3-34.4) | 44 / 0 | 1.5 / 3.0T | T2 | 0.5 | NesVor |
| | ZRH synth T2** | 3 | 30 | 33.9 ± 0.8 (32.9-34.8) | 12 / 18 | Synthetic | T2** | 0.5 | GAN (label to image) |
| Test: out-of-domain | dHCP pretT1 | 2 | 40 | 33.4 ± 2.3 (29.3-40.9) | 0 / 40 | 3.0T | T1 | 0.5 | SVRTK |
| | CHUV | 4 | 40 | 28.4 ± 4.2 (21.0-35.0) | 25 / 15 | 3.0T | T2 | 1.125 | MIAL-SR |
| | ZRH synth T1** | 3 | 30 | 33.9 ± 0.8 (32.9-34.8) | 12 / 18 | Synthetic | T1** | 0.5 | CycleGAN (image to image) |
| | dHCP pretT2 | 2 | 40 | 33.4 ± 2.3 (29.3-40.9) | 0 / 40 | 3.0T | T2 | 0.5 | SVRTK |
| | KCL | 5 | 20 | 31.0 ± 3.5 (23.0-36.3) | 15 / 5 | 0.55T | T2 | 0.5 | SVRTK |
| | Vienna | 6 | 40 | 24.9 ± 4.8 (18.1-350) | 20 / 20 | 1.5 / 3.0T | T2 | 1.0 | NIFTYMIC |
| | ZRH Spina | 3 | 90 | 27.3 ± 1.2 (21.1-29.7) | 0 / 90 | 1.5 / 3.0T | T2 | 0.5 | NesVor |
| | **Total** | | **759** | **28.9 ± 4.1 (18.1-40.9)** | **439 / 320** | | | | |

*Data sources: 1: FeTA2021 Dataset; 2: developing human connectome project (dHCP); 3: University Children's Hospital Zürich, 4: Centre hospitalier universitaire vaudois (CHUV); 5: King's College London; 6: Medical University of Vienna

**: Synthetic images

***: Gestational weeks or corrected gestational age.

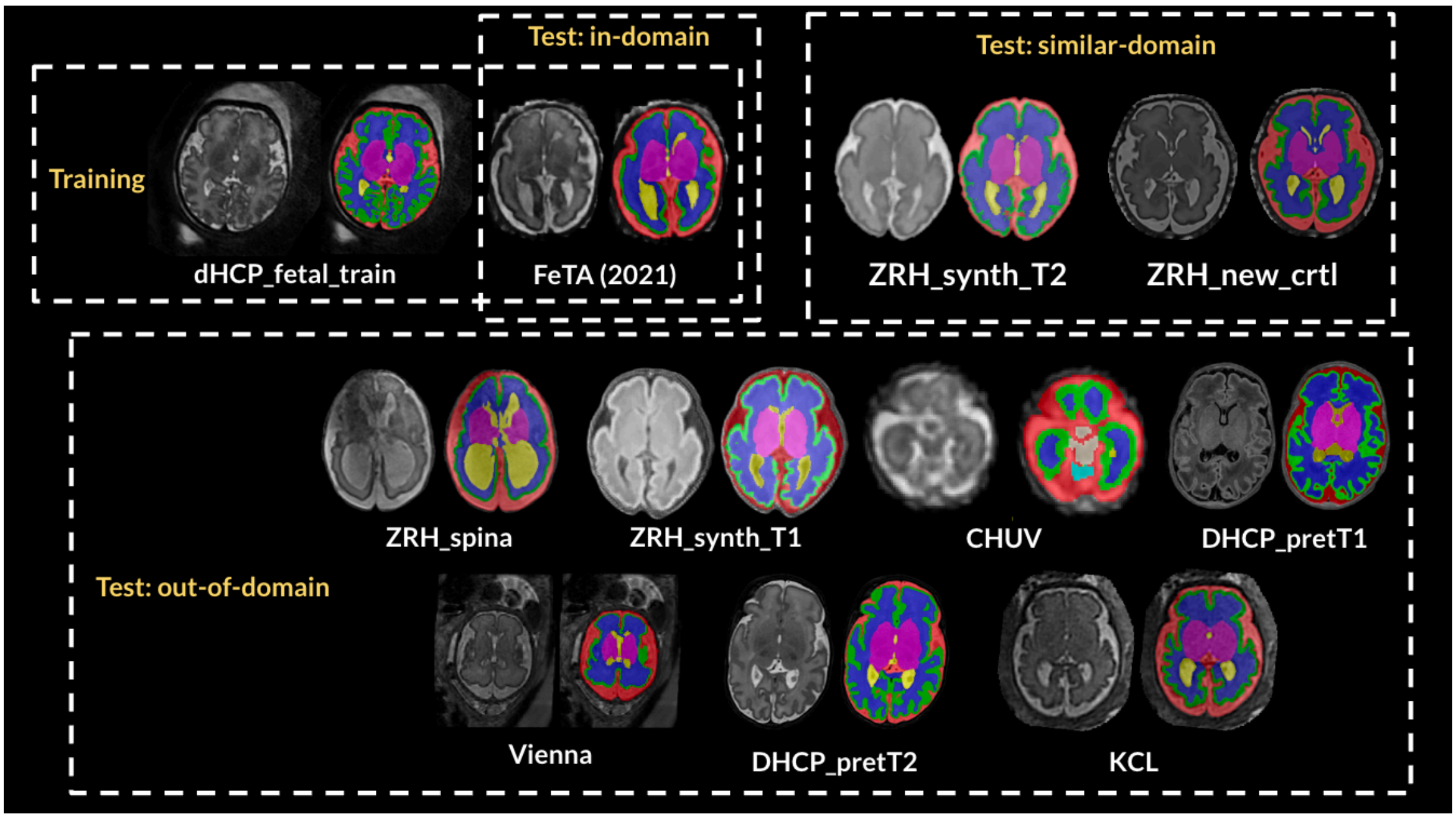

Figure 1: The training/testing datasets and their grouping.
Visualizations of MRI datasets and corresponding segmentation templates used in our study. This image shows the two training datasets and three groups of testing datasets (in-domain, similar-domain, and out-of-domain).

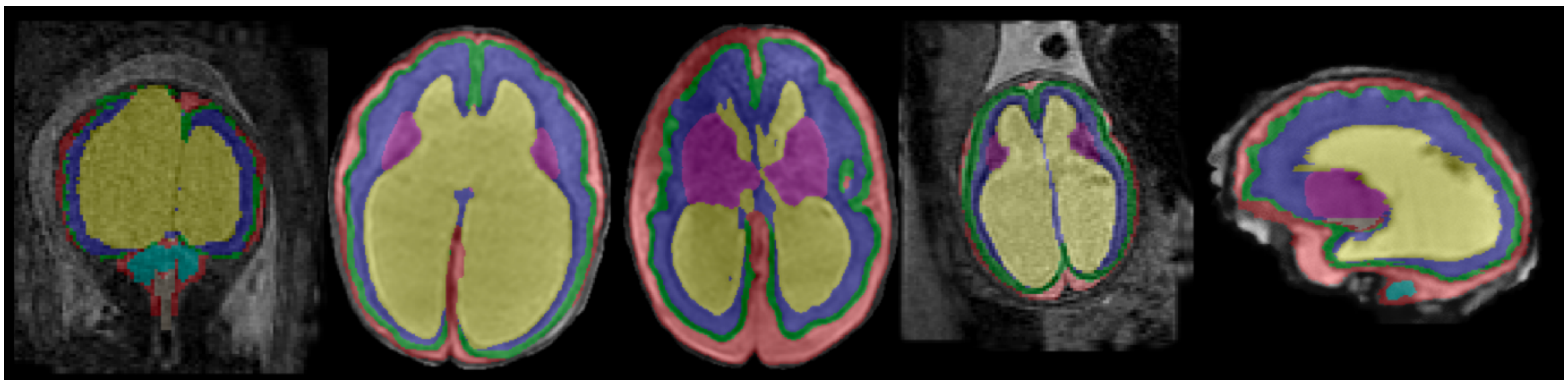

Figure 2: Examples of templates from the "most-pathological" test set. The images show various degrees of moderate (3rd image) to severe (rest of the images) ventriculomegaly as a consequence of neural tube defects, such as open spina bifida.

### 2.1.2 Anatomical structure maps

Manual expert segmentations according to the FeTA Challenge's anatomical nomenclature were provided for the FeTA, ZRH synth CHUV, KCL and Vienna datasets. A human-in-the-loop method was utilized for segmenting brains in the spina bifida ("ZRH Spina") dataset: these cases were segmented using an nnU-Net (Isensee et al., 2021) based segmentation network trained on the FeTA dataset, corrected for potential errors, and re-trained using the corrected training data. The MR images of dataset "ZRH new crtl" were segmented using the BOUNTI (Uus et al., 2023), visually checked for segmentation errors, manually corrected if necessary, and a script was used that converted the BOUNTI segmentation templates to the FeTA segmentation standard. The dHCP data was released with segmentation templates. We used an in-house script to convert the dHCP templates to FeTA segmentation standard.

### 2.2 Data-driven sampling for training

Due to the limited diversity in training data for fetal MR image segmentation networks, there is a risk of sampling imbalance when pooling from a training set, which could lead to reduced generalizability of the resulting networks (Zhang et al., 2020). Previous works addressing such training imbalance have focused on data augmentation (Zhang et al., 2020), harmonization (Dinsdale et al., 2021; Wachinger et al., 2021), and balanced sampling (Meier et al., 2019; Rebsamen et al., 2019). Focusing on balanced sampling, our data-driven pipeline is able to automatically divide any training set into subgroups sharing similar 3D morphological characteristics without the need of any meta-information regarding the subjects. Then, we assigned different sampling weights to templates in each subgroup to ensure balanced representation across subgroups during training. The pipeline, demonstrated in Figure 3, is divided into three steps:

First, from each training template, we extracted a total of 21 shape features to characterize the morphology of various brain structures. The 21 features were then concatenated into a characteristic vector. The contents within each characteristic vector are demonstrated in Figure 4. The vector consists of two components, structural shape representations and global shape representations. For structural shape representations, we selected six structures from each training template (white matter, gray matter, ventricles, cerebellum, thalamus/deep gray matter, brainstem). Extra-axial CSF is excluded as it is not a stable and coherent anatomical structure. For each structure, we calculated three values: its volume divided by the total volume of all structures, its surface area divided by the surface area sum of all structures, and its surface area to volume ratio. This results in a total of 3*6=18 structural features for each brain template. For global shape representations, we used the whole mask of all the six structures combined to calculate its total volume, total surface area, and surface area to volume ratio. The three global representation features are concatenated with the 18 structural representations to form an overall characteristic vector of 21 features for each training brain template.

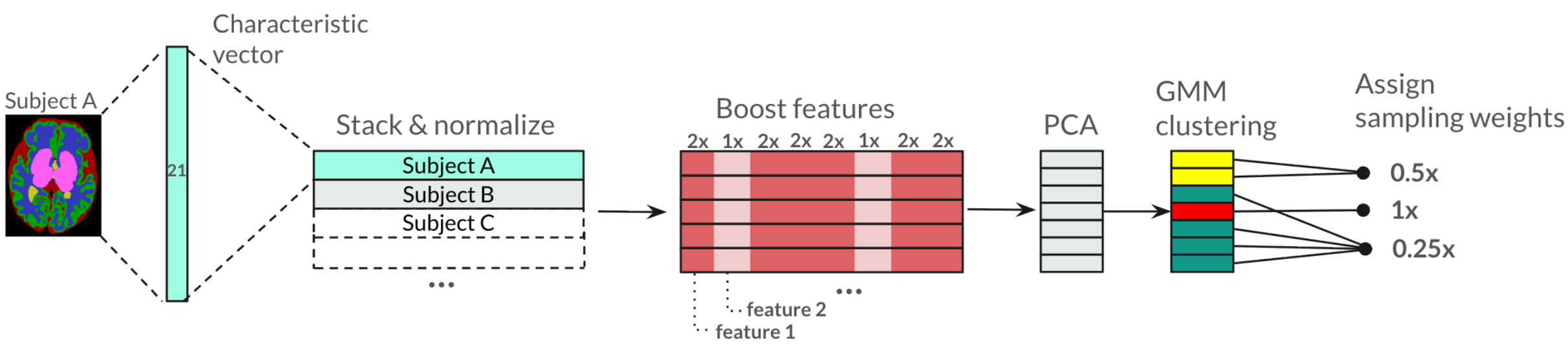

Figure 3: Pipeline for data-driven sampling
We extract shape information (i.e. the characteristic vector) from each training annotation template, process them, and divide the whole training set into multiple groups by clustering the characteristic vectors. The groups are used to assign sampling weights for every training template, where each group has the same possibility of being chosen despite the number of templates within. This increases the training exposure to under- represented templates and decreases the training exposure to over-represented templates.

Then, we processed the characteristic vectors and divided them into separate groups: after stacking the vectors together, we normalized each feature to 0-1 using min-max scaling. Based on the assumption that certain features are more important in determining the shape characteristics of a template, we assigned a 2x weighting boost to all global representation features and the subset of structural shape representation features corresponding to larger or more complex structures, including the cortex, white matter, and lateral ventricles. The features that are boosted are shown in dark red in Figure 4. We expect linear relationships among the extracted features as they are mostly defined as ratios, either between multiple attributes of the same structure or between the same attributes of multiple structures. Thus, we applied Principal Component Analysis (PCA) to reduce the dimensionality of all features to the first three principal components. These low-dimension features were then divided into a certain number (determined by examining the dataset size and its shape disparity) of subgroups by applying Gaussian Mixture Model (GMM)-based clustering.

Finally, to ensure that each subgroup is equally represented during training, our method assigned different sampling weights to each template so that every subgroup is selected with the same probability - regardless of how many templates it contains - and that all templates within a given subgroup are equally likely to be chosen. Thus, the sampling weight of a template within a subgroup **G** will be assigned according to Equation 1. In this case, a template coming from a group with a small number of templates (i.e., possessing underrepresented shape characteristics) would have a relatively larger sampling weight.

$$\text{weight}(\mathbf{G}) = \frac{1.0}{(total\ \#subgroups) * (\#templates\ within\ group\ \boldsymbol{G})} \tag{1}$$

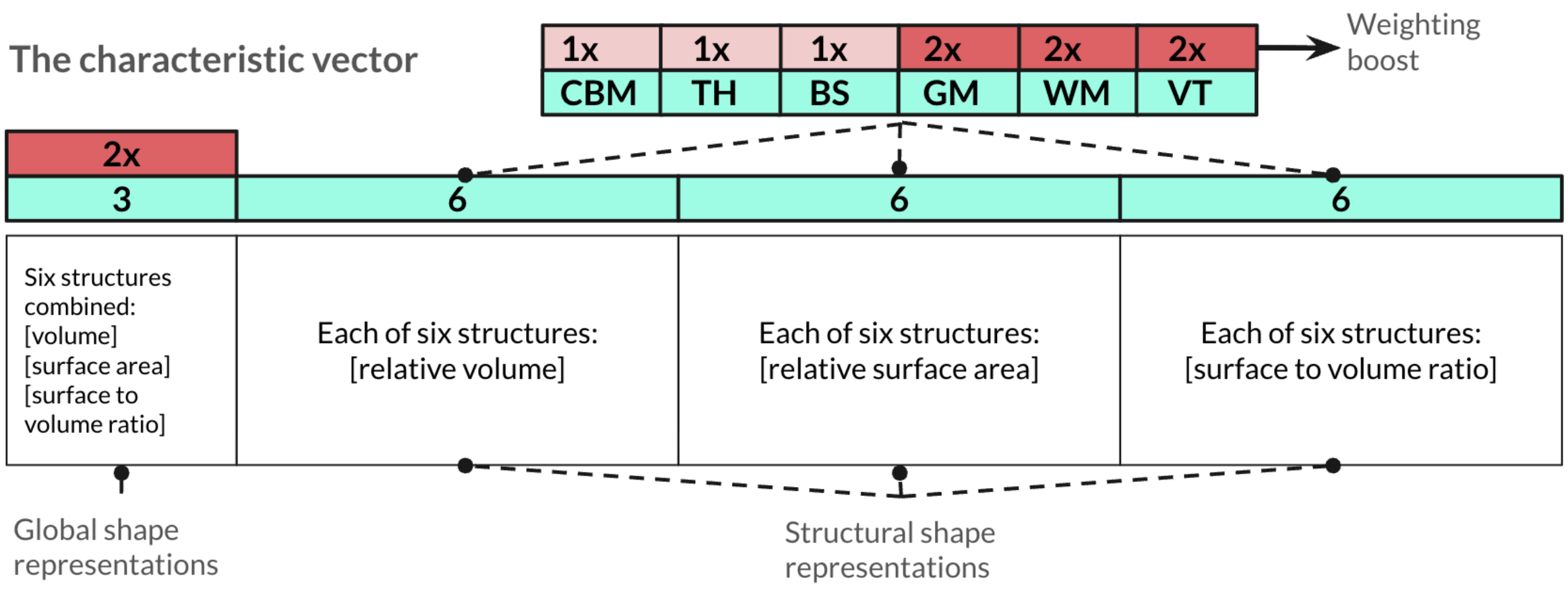

Figure 4: The characteristic vectors
The content within the characteristic vectors extracted from every training template. Each vector consists of three global representation features and 18 structural representation features, where three values are extracted for each of the six shown structures. Among the 21 features, 12 of them are given a 2x weighting boost. CBM: cerebellum, TH: thalamus/deep gray matter, BS: brainstem, GM: gray matter, WM: white matter, VT: ventricle.

## 2.3 Generation of pathological training templates for shape-based domain shifts

To simulate ventriculomegaly and hydrocephalus (expanded ventricles), we created one additional template for each original FeTA training template through the structural deformations introduced in (Kaandorp et al., 2025). First, we defined the maximum dilation to be when the ventricles cover 65% of the white matter. Then, for each hemisphere of each template, a random number of dilations below this maximum was selected to determine how much the ventricles would be expanded. The ventricles were dilated into the surrounding white matter label and smoothed while also enforcing the boundary of the resulting ventricles to be at least two voxels apart from any other structure to avoid unrealistic overlaps. Examples of subjects with dilated ventricles can be found in visualized pipeline in Figure 5. Ventriculomegaly and hydrocephalus are chosen here for two main reasons. First, those two pathologies result in significant structural shape variations, which is the main issue we aim to address. Second, the two pathologies are typical in fetuses and are prevalent in both the training and testing datasets we use. Fetuses with these conditions are commonly referred by physicians for further evaluation on fetal MRI, and automated segmentation of cerebral structures may provide valuable added information to this process.

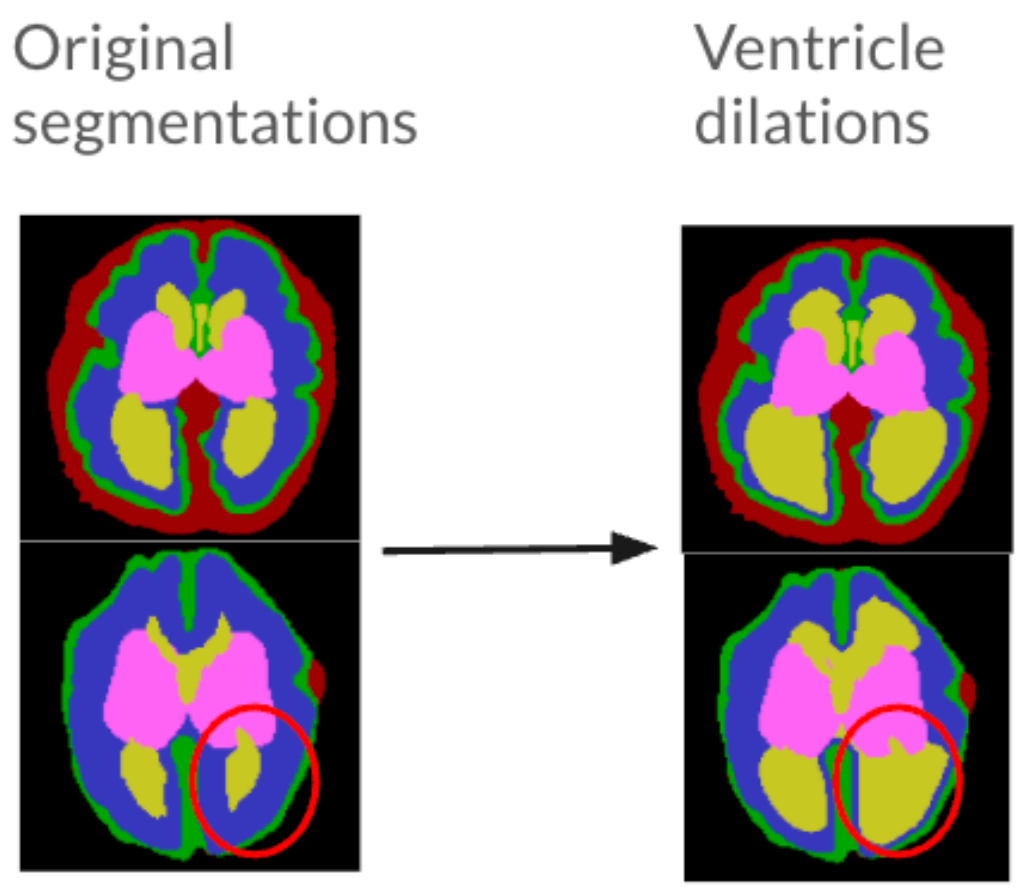

Figure 5: Examples of generated pathological training templates
We generate one pathological template for each training subject using random dilation parameters. This new generated set is combined with the original training set and fed into our data-driven sampler and subsequently SynthSeg. The circled regions on the ventricles highlight the differences between the original and synthetic pathological training templates.

## 2.4 Learning background

As fetal MR images include the amniotic fluid and maternal organs, the generated images may contain many structures beyond the brain. Our datasets included mixed post-processing, meaning partially skull-stripped data and data with surrounding organs. Thus, for the dHCP fetal training data, we created four additional labels (i.e. training structure) representing the surrounding structures by adopting the structural subdivision in (Zalevskyi et al., 2024). Specifically, we recorded the intensities for every artifact voxel (i.e. possessing a non-zero intensity) in the background, scaled them to 0-1, and divided them into four groups using k-means clustering. The voxel positions within each group were assigned the same label in the training template. Each additional label was treated as a distinct structure when SynthSeg generates the training images but converted back to the background label when SynthSeg generates the training ground truth segmentations. For the FeTA training cases, which often contain fewer artifacts, we assigned all artifact voxels to a single additional label. In this way, the network would be able to learn about both structurally diverse and homogeneous backgrounds.

## 2.5 Train/test specifications and model ensembling

In the training templates, because the brain's position can vary within the volume, we first cropped the brain volume to its minimum bounding box. Then, we padded the background voxels equally along each dimension. In this case, the training brains would be in the center of each volume.

For every training run, the network was trained with a learning rate of 1e-4 using an Adam optimizer for 20 epochs, with 5000 iterations per epoch and a batch size of 1. A checkpoint was saved for each epoch, and by evaluating every checkpoint starting from epoch 12 (after 60% of the training time) using the corresponding MR images of the original FeTA training templates, we selected the best checkpoint to be applied to the testing datasets.

At inference, the testing datasets underwent several post-processing steps to ensure uniformity in their appearance. First, to eliminate excessive amounts of redundant black backgrounds, we cropped each input image by the minimum bounding box of voxels that have a non-zero image intensity and pad only a small margin of 5 voxels of background around the cropped volume. Then, for the testing templates with voxel sizes of 1 or larger (i.e., a resolution much lower than the 0.5 $mm^3$ training templates), we first upsampled the whole volume to 0.6$mm^3$ resolution, applied the network to the upsampled version, then downsampled the resulting segmentations to its original affine space. If the volume had a voxel size close to 0.5$mm^3$, we directly applied the trained network to generate the segmentations.

As we observed that the network results may fluctuate due to randomness in the training process (i.e., a difference in network accuracy between two runs of the identical training script), for every experiment reported, the training script was run three times. The final segmentation was generated by merging the outputs of the three networks via model ensembling, implemented using max posterior model selection: when networks disagree on a voxel, the network with the highest posterior probability (the most confident) decides the final result. This gives an optimistic assessment of the capabilities of each method, which is used to compare between each of our experiments. In Section 4.3, when comparing with other related works that do not employ emsembling, for a fairer comparison, we provide additional results where we no longer perform output merging and instead use the average metrics of the three networks individually evaluated.

## 2.6 Summary of the model evaluation

We conducted eight main experiments designed to systematically assess the efficacy of this synthetic approach and the effects of training with different datasets and sampling strategies. A strategy to vary the different training datasets, sampling methods, and use of synthetically generated training templates has been employed, as described below, with a summary shown in Table 2.

**Experiment 1 (baseline):** *Training data*: FeTA_train. *Sampling*: uniform sampling. (This unmodified implementation of SynthSeg serves as the baseline)
**Experiment 2 (samp):** *Training data*: FeTA_train. *Sampling*: data-driven sampling, where all training templates are divided into four groups.
**Experiment 3 (dhcp):** *Training data*: FeTA_train and dHCP_fetal. *Sampling:* uniform.
**Experiment 4 (dhcp+samp):** *Training data*: FeTA_train and dHCP_fetal. *Sampling:* we use Data-driven sampling independently on each dataset, where the FeTA_train set is divided into 4 groups and the dHCP_fetal set is divided into 8 groups.
**Experiment 5 (synth):** *Training data:* FeTA_train and synthetic templates. *Sampling:* uniform.
**Experiment 6 (synth+samp):** *Training data:* FeTA_train and synthetic templates. *Sampling:* Data-driven sampling, where the whole training set is divided into 6 clusters.
**Experiment 7 (dhcp+synth):** *Training data:* FeTA_train, synthetic templates, and dHCP_fetal. *Sampling:* uniform.
**Experiment 8 (dhcp+synth+samp):** *Training data:* FeTA_train, synthetic templates, and dHCP_fetal. *Sampling:* data-driven sampling is applied independently to two subsets-- Templates from [FeTA_train and synthetic templates] are divided into 6 groups and templates from dHCP_fetal are divided into 8 groups.

For all experiments using the dHCP_fetal training set for training, in addition to the aforementioned sampling methods, we added a constraint where the sum of the sampling weights of all templates in dHCP_fetal always adds up to 50% so that difference in subject numbers between the various training sets does not serve as a confounder. All processing (i.e., normalization, boosting, PCA, clustering, grouping, and weight assignments) of the extracted characteristic vectors from dHCP_fetal are conducted separately from the rest of the training set (i.e., FeTA_train and the synthetic templates). All processing for FeTA_train and the synthetic pathological templates are conducted together as they both originate from FeTA_train.

The rationale behind the eight experiments are three main questions we aim to answer: First, what does data-driven sampling achieve? Second, what does the generated pathological training templates achieve? Third, as our methods act upon the training datasets, how does changes in the training dataset composition affect the outcomes? As listed in columns 2, 3, and 4 in Table 2, the experiments form a permutation of including/excluding the respective component corresponding to each of the three major questions, forming a total count of eight experiments.

To assess our network performances, we calculated the Dice score between the ground truth and generated segmentations for a subject by averaging the Dice scores for all structures excluding the background. Paired t-tests are conducted on the scores to evaluate the significance of the improvements of our methods. For the most-pathological dataset, since the number of data points are not very large (n=33) and the improvements discussed in Section 3 did not pass the Shapiro-Wilk normality test, the Wilcoxon test are conducted as a supplementary to the paired t-tests. We also qualitatively evaluated our methods through visual assessments.

Table 2: Our eight experiments

There are mainly three components that could be varied during the training process:, 1, (on top of FeTA_train) whether to include an additional dHCP_fetal for training; 2, whether to use synthetic pathological templates; 3, whether to employ data-driven sampling. This is an overall summary of the methodology for every experiment.

| Experiment | Train set: FeTA | Train set: dHCP | Synthetic templates | Data-driven sampling |
|---|---|---|---|---|
| Experiment 1: baseline | ✓ | | | |
| Experiment 2: samp | ✓ | | | ✓ |
| Experiment 3: dhcp | ✓ | ✓ | | |
| Experiment 4: dhcp+samp | ✓ | ✓ | | ✓ |
| Experiment 5: synth | ✓ | | ✓ | |
| Experiment 6: synth+samp | ✓ | | ✓ | ✓ |
| Experiment 7: dhcp+synth | ✓ | ✓ | ✓ | |
| Experiment 8: dhcp+synth+samp | ✓ | ✓ | ✓ | ✓ |

## 2.7 Ethics statement

The corresponding local ethics committees independently approved the studies under which data were collected, and all participants gave written informed consent. The overall study protocol and use of the shared data was approved by the Cantonal Ethical Committee of Zürich (Decision number 2022-01157). The Viennese cohort data was acquired as part of a retrospective single-center study and was anonymized and approved by the ethics review board and data clearing department at the Medical University of Vienna, responsible for validating data privacy and sharing regulation compliance (Ek Nr. 2199/2017 and 1585/2021). This CHUV dataset was part of a larger research protocol approved by the ethics committee of the Canton de Vaud (decision number CER-VD 2021-00124) for re-use of their data for research purposes and approval for the release of an anonymous dataset for reproducible research and open science purposes. The KCL low field data was acquired as part of the MEERKAT study, approved by the local ethics committee (Bromley Research Ethics Committee 21/LO/0742, 8/12/2021). The informed consent process explicitly included approval for sharing with academic partners.

## 3. Results

The results are congregated into four domains according to Section 2.1.1. Figure 6 visualizes the Dice score (averaged across all structures) distributions of all subjects, separated by domains. Figure 7 shows the Dice score distributions of each individual structure, separated by domains. Table 3 shows the mean and standard deviation of the Dice scores (average of all structures) for each domain. The 95th percentile Hausdorff distance (HD95) distributions for each domain could be found in Figure 8.

The average Dice score of each structure among all subjects in each domain could be found in Supplementary Tables 1, 2, 3, and 4. An additional distribution plot of the Volume Similarities (VS) grouped by domain could be found in Supplementary Figure 1. In addition, we separate all underlying datasets that compose each test domain and report their results separately for every experiment in Supplementary Table 5.

### 3.1 In-domain

For the in-domain data, all eight experiments have similar results across all structures, with the baseline having the largest Dice with a slight margin as per table 3 and experiment 8 having the smallest Dice (0.03 smaller than the baseline). We also observed that in experiments 3,4,7, and 8, where the training set contains the dHCP training templates (with larger developmental ages), the models perform slightly better in the cortical gray matter.

### 3.2 Similar-domain

For the similar-domain data, Table 3 shows that in experiments trained with dHCP (experiments 3,4,7,8), subjects generally had slightly better results compared to their counterparts without dHCP (experiments 1,2,5,6), respectively having average Dice improvements of 0.02, 0.04, 0.04, and 0.03 (all with $p<1e-9$). From Figure 7, we also observed that the addition of data-driven sampling has caused a decrease in the performance of the ventricles when the training data did not contain the synthetic templates (experiments 1,2,3,4), but an increase in performance when the training data contained the synthetic template but not the dHCP training set (experiments 5 and 6).

### 3.3 Out-of-domain

For the out-of-domain data, all the observations from the similar-domain sets are retained, except that the changes in ventricle scores caused by the data-driven sampling became much smaller. According to Table 3, the only noticeable increase in the overall mean Dice scores as a result of adding the dHCP occurred when comparing experiments 5 and 7 (improvement of 0.02, $p<1e-8$). The overall result disparity between each experiment is also much smaller compared to the similar-domain test set.

### 3.4 Effects of pathology on model outcomes

When tested on the most-pathological domain, the performances of all models degraded significantly compared to the other domains, which is expected as this domain contains the most significant shape variations. Comparing the results across experiments tested on this domain most clearly highlights the effects of our methods. We observed from Table 3 that the data-driven sampling has caused a notable improvement when the training data contains only the FeTA training set (experiments 1 and 2), with an average Dice score improvement of 0.07 ($p<1e-4$ for t-test and $p<1e-5$ for Wilcoxon test). The data-driven sampling also slightly improved the average Dice scores when the training data contained FeTA+dHCP (0.03 improvement from experiment 3 to 4, $p=0.07$ for t-test and $p=0.4$ for Wilcoxon test, not statistically significant) or FeTA+synthetic (0.02 improvement from experiment 5 to 6, $p < 0.01$ for t-test and $p<1e-5$ for Wilcoxon test). The only decrease occurred when the training data contained all of FeTA+dHCP+synthetic (0.03 decrease from experiment 7 to 8). There is also a clear trend that including dHCP during training (experiments 3,4,7,8) lowered the overall results compared to their counterparts (experiments 1,2,5,6), respectively having average Dice decreases of 0.08, 0.12, 0.08, and 0.13.

From Figure 7, we see that the differences between experiments are most extreme in the lateral ventricles, which is where most of the pathology-related deformations take place. We found that the usage of data-driven sampling significantly improved the ventricle segmentations when the training data did not contain the dHCP set (experiments 1,2,3,4), though it also caused a slighter decrease in segmentation performance when the training data included the synthetic templates (experiments 5,6,7,8). We also note that, compared to the experiment 1 baseline, the inclusion of synthetic training templates greatly improved the ventricle segmentations when the training data did not contain dHCP templates (experiments 5 and 6).

Without data-driven sampling, we observe that the model performances correlates to how good the training and testing datasets match in terms of the relative ratio of pathological subjects, where similar compositions result in better outcomes. For experiments 1, 3, 5, 7 (all without data-driven sampling), the effective percentage (please refer to Supplementary Note 1) of regular templates in each training set are, respectively, 39%, 69%, 19%, and 60%, with exp3 > exp7 > exp1 > exp5. As shown in Table 3, this is the exact same order of their performances ranked on the similar-domain (the domain with the smallest percentage of pathological sets) from best to worse and the exact opposite order of their performances ranked on the most-pathological domain (100% pathological). This indicates that a model trained with higher percentages of pathological cases performs better on a test set with a high pathological ratio alike, while a model trained with lower pathological percentages performs better on a test set with a low pathological ratio. However, if we take the data-driven-sampled counterpart of those experiments (i.e., in order--experiments 2, 4, 6 and 8), the performances ranked on similar-domain becomes exp8 = exp4 > exp6 > exp2 and the performance on most-pathological becomes exp2 > exp6 > exp4 > exp8. Comparing this with the percentage of regular templates used for training (exp4 > exp8 > exp2 > exp6) we see that data-driven sampling has clearly weakened the relationships between model performance and how well the training/testing sets match in their regular/pathological compositions.

### 3.5 Performances in distance-based matrices

In addition to the Dice scores, we report the 95th percentile Hausdorff distances (HD95) in Figure 8 to assess whether the trends observed in the Dice scores are reflected in boundary qualities. Generally, HD95 performances correspond well to the performances measured in Dice scores.

For the in-domain subjects, just as the Dice scores, all experiments had similar performances, with the only difference being experiments 5 and 6 performing slightly worse compared to their relative performance in Dice scores. For similar-domain, the box positions of experiments 3,4,7, and 8 are better than the other experiments. This is consistent with their Dice scores. However, if we look at the average, we observe the opposite, where those four experiments performed slightly worse compared to the other experiments. This indicates that such experiments are affected by worse-performing outliers. Upon further inspection, we found that the outliers likely origin from a subset of ZRH_NEW_crtl, which contains a series of low-quality MR images. The models of those experiments tend to produce many holes and islands in their segmentations, which increases their HD95 mean but not their medians and interquartile ranges.

For the out-of-domain subjects, just as their Dice counterparts, the HD95 performances for all experiments are similar, with experiments 2, 5, and 6 only performing marginally worse. The results for the most-pathological domain also corresponds well to the Dice distributions, with experiments 1,2,5, and 6 (all experiments trained without dHCP_train) generally outperforming the other experiments and the data-driven sampling significantly improving the results of experiments from 1 to 2, 3 to 4, and 5 to 6, while decreasing the performance from experiment 7 to 8.

### 3.6 Qualitative assessments

We also performed qualitative visual assessments of the segmentation quality, conducted through a collaborative effort of the authors by comparing the network outputs with the ground truths. We provide examples of subjects from the in-domain (Figure 8), similar-domain (Figure 9), and out-of-domain datasets (Figure 10) covering images with different amounts of shape variance, background artifacts, and resolutions. In the "*Vienna*" dataset in Figure 10, we observe that in experiments where templates with an increased amount of background artifacts (i.e., dHCP_fetal) are included and processed with background subdivision (experiments 3, 4, 7, 8), over-segmentations into the background did not occur despite such over-segmentations taking place in all other experiments.

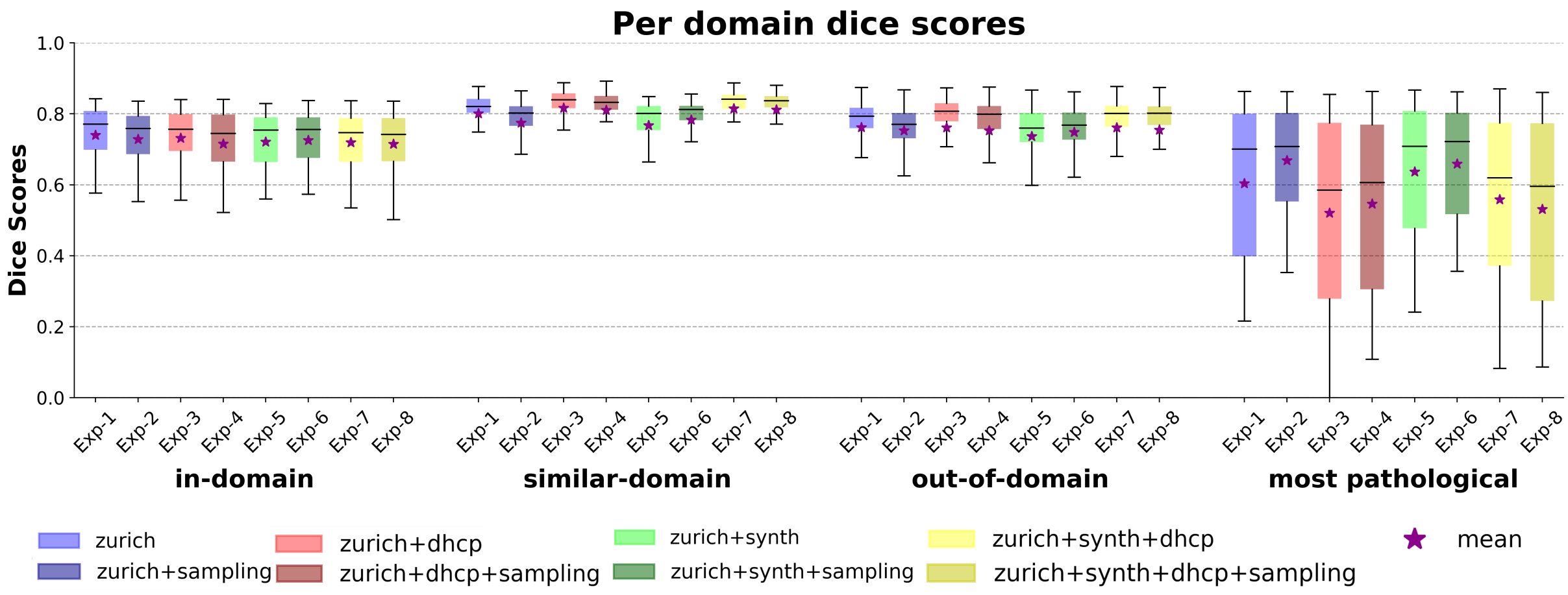

Figure 6: Comparison of results between domains
For each domain, we plot the distribution of the Dice (mean of all structures) score for every subject. As shown in the legend, the eight experiments are grouped into four color-matched pairs. For each pair, the brighter color corresponds to the experiment without data-driven sampling, while the darker color corresponds to the same experiment but with our data-driven sampling added.

Table 3: Dice scores result summary.
Dice scores among all testing subjects in each domain, all structures averaged for each subject (mean ± standard deviation).

| Experiment | in-domain | similar-domain | out-of-domain | most-pathological |
|---|---|---|---|---|
| Experiment 1: baseline | **0.74**±0.09 | 0.80±0.11 | **0.76**±0.11 | 0.60±0.21 |
| Experiment 2: samp | 0.73±0.09 | 0.77±0.11 | 0.75±0.08 | **0.67**±0.15 |
| Experiment 3: dhcp | 0.73±0.11 | **0.82**±0.12 | **0.76**±0.15 | 0.52±0.26 |
| Experiment 4: dhcp+samp | 0.72±0.12 | 0.81±0.12 | 0.75±0.14 | 0.55±0.23 |
| Experiment 5: synth | 0.72±0.09 | 0.77±0.12 | 0.74±0.10 | 0.64±0.19 |
| Experiment 6: synth+samp | 0.73±0.09 | 0.78±0.10 | 0.75±0.09 | 0.66±0.16 |
| Experiment 7: dhcp+synth | 0.72±0.10 | 0.81±0.12 | **0.76**±0.13 | 0.56±0.23 |
| Experiment 8: dhcp+synth+samp | 0.71±0.11 | 0.81±0.12 | 0.75±0.14 | 0.53±0.25 |

## 4 Discussion

Our study aimed to address key limitations in automated fetal MR image segmentation related to both contrast- and shape-based domain shifts in pathological cases. We proposed a data-driven template sampling strategy to enhance the anatomical diversity of training data, leveraging shape representations to optimize template selection. By integrating this sampling approach with augmentation techniques, we aimed to improve the generalizability of CNN-based segmentation models to real-world clinical data including cases with severe hydrocephalus/ventriculomegaly. Our methods are evaluated on a diverse collection of test sets via a comprehensive component analysis broken down into eight experiments. Our findings demonstrate that our strategy improved the segmentation for cases with substantial pathologies, though it also comes with a smaller decrease in performance for neurotypical subjects. The addition of ventriculomegaly training templates also achieved a similar effect, where the network is oriented towards recognizing pathological cases at the cost of its precision in neurotypical structures. Experiments including the dHCP training set and background subdivision resulted in an improved ability to discern between the brain and background artifacts as well as better segmentation performance in the similar-domain and out-of-domain datasets but comes with a larger performance drop in cases with severe anatomical variations.

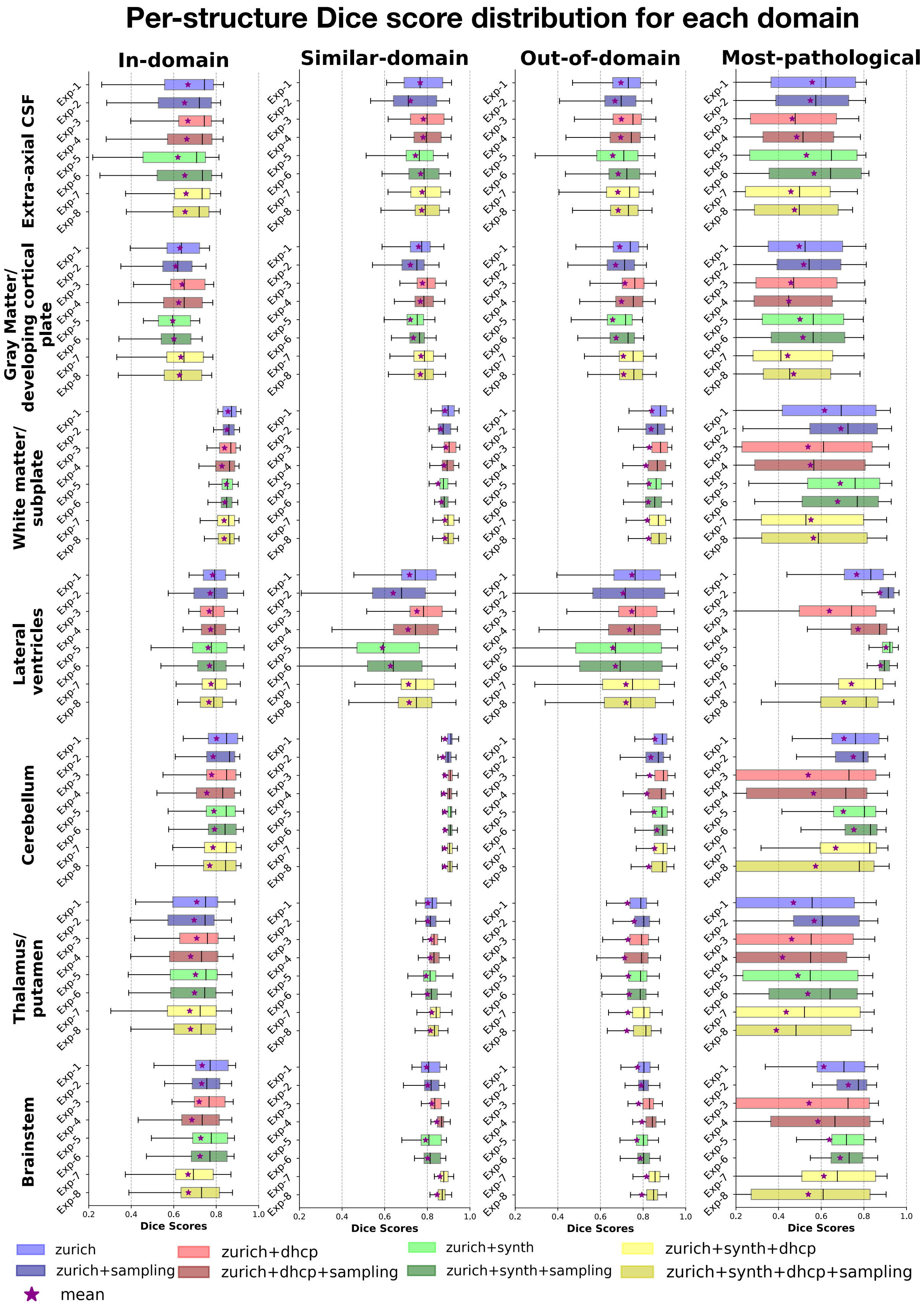

Figure 7: Per-structure results separated by domain.
For each of the four domains (separated into four columns), we plot the Dice score distribution of each brain structure for each experiment. As shown in the legend, the eight experiments are grouped into four color-matched pairs. For each pair, the brighter color corresponds to the experiment without data-driven sampling, while the darker color corresponds to the same experiment but with our data-driven sampling added.

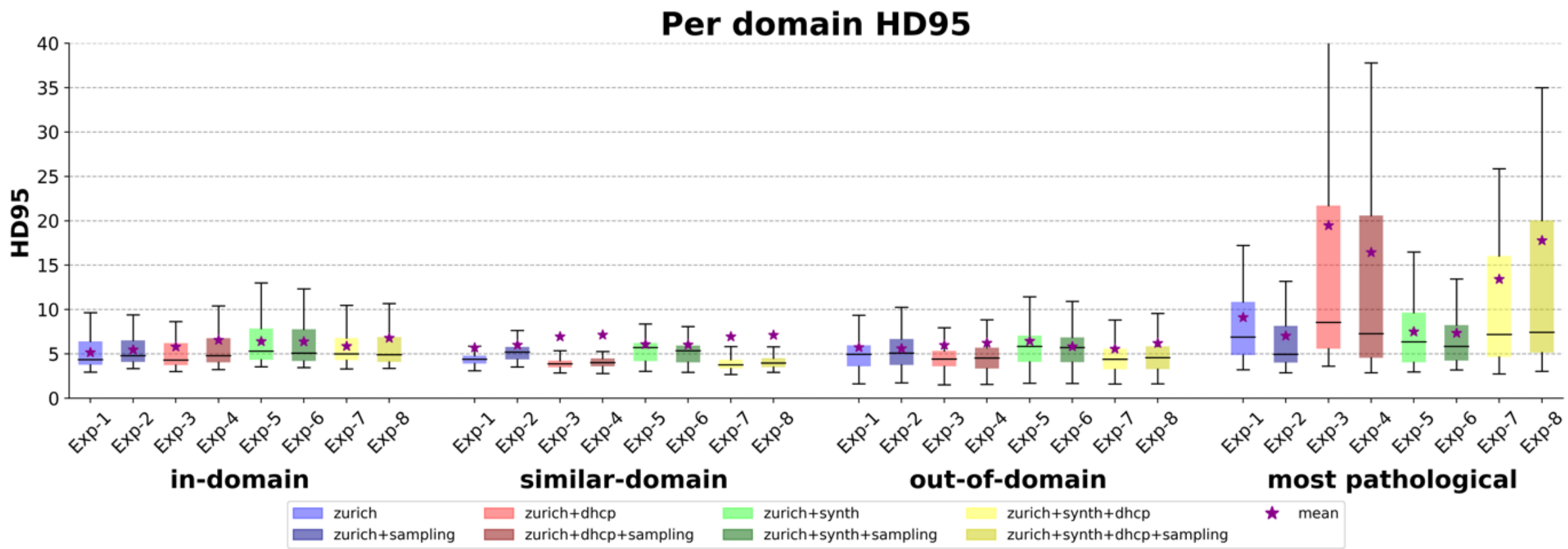

Figure 8: Comparison of the 95th percentile Hausdorff distances (HD95) between domains.

For each domain, we plot the distribution of the HD95 mean of all structures for every subject.

## 4.1 Effects of data-driven sampling

Overall, the data-driven sampling did not significantly affect the results on any structure in the in-domain testing set. This indicates that oversampling templates with less-represented shape characteristics with training does not affect the results when the domain shift remains minimal.

However, in the similar-domain set, we do see a more noticeable decrease in performance for the ventricles when the training data lacked synthetic templates (experiments 1,2,3,4). This could be attributed to the absence of pathological templates in the similar-domain test set, where networks trained with either synthetic templates or oversampled minorities may be susceptible to ventricle over-sampling on this test set (demonstrated in Figure 9, experiments 2,5,6). In contrast, the opposite happened when the training data contained synthetic templates (experiments 5 and 6) where the ventricle segmentation improved after applying data-driven sampling. We believe this is due to the abundance of training segmentation templates with inflated ventricles introduced by the addition of one synthetic ventriculomegaly map for each original FeTA training template, which caused this feature to be a commonality instead of a minority. This means templates with this feature are no longer likely to be further oversampled under the influence of the data-driven sampling. These observed effects of data-driven sampling could also be found within the out-of-domain test set, but the performance differences were greatly reduced. Presumably, this was caused due to the inclusion of pathological templates within this large and diverse set.

In the most-pathological testing set, the opposite of the aforementioned effects was observed in experiments 1,2,3,4 when considering the segmentation quality of the ventricles, where the data-driven sampling improved the segmentations. Moreover, such improvements are much more substantial compared to any other testing set. The data-driven sampling also caused a decrease in the ventricle segmentation quality for experiments where the synthetic templates were used. This further demonstrates that the synthetic templates already exposed the network to dilated ventricles sufficiently during training, so data-driven sampling did not add any extra emphasis to this feature. It is possible that the emphasis was instead moved to other properties, given that the performance improved in most of the other structures from experiment 5 to 6.

It is also important to note that the data-driven sampler is not specific to SynthSeg, where the sampling scheme could be easily adopted to other supervised pipelines for training MR image segmentation networks.

## 4.2 Effects of additional training data

### 4.2.1 The synthetic ventriculomegaly set

Excluding the most-pathological testing set, adding the synthetic training templates (experiments 5,6,7,8) generally did not cause notable changes compared to their counterparts (experiments 1,2,3,4), with the only exception being comparing experiments 1 and 5, where decreases in performance after adding the additional dataset were observed. However, in the most-pathological dataset, experiment 5 achieved a 0.04 higher average Dice score compared to experiment 1, indicating that increasing pathology representation by adding more templates with inflated ventricles tends to skew the network towards recognizing inflated ventricles at the expense of the ability to identifying normal ventricles. However, when combined with data-driven sampling (comparing experiments 2 and 6), the discrepancies in segmentation performances were greatly reduced, illustrating the ability of the data-driven sampling to maintain representation balance within a diverse training dataset.

### 4.2.2 The dHCP training set

Overall, the inclusion of dHCP templates improved the results for the cortex and ventricles for the similar-domain and out-of-domain sets and for the cortex on the in-domain set. This shows that SynthSeg benefits from learning from larger variability of structures, such as the cerebral cortex, more precise segmentations, and smaller ventricles (the dHCP training templates included fetuses in later gestation compared to the FeTA 2021 dataset). However, the opposite was observed in the most-pathological test set, where the scores for CSF, white matter, gray matter, and ventricles were lower compared to their counterparts. This shows that the network was not able to generalize well to subjects with very exaggerated shape differences, where the large presence of templates with older age and smaller ventricles likely occluded the representation of existing pathological templates.

Our experiments also validate the usage of training templates generated using background subdivision on MR images with large amounts of background artifacts. This is demonstrated in the second row of Figure 11, where all experiments in which background subdivision was not used (experiments 1,2,5,6) contain over-segmentations into the background, whereas this problem did not occur in any of the experiments utilizing the dHCP training set with background subdivision (experiments 3,4,7,8).

## 4.3 Comparisons with more related works

In addition to the main baseline of this work--the original SynthSeg (Billot et al., 2023) shown as Experiment 1--it would be interesting to compare it with more related works. Although our multi-centric settings are not common in existing papers due to its large diversity in both training and testing datasets, several works related to domain shift and fetal brain segmentation has been identified for comparison, including FetalSynthSeg (Zalevskyi et al., 2024), the FeTA challenge series (Payette et al., 2025, 2023, 2021; Zalevskyi et al., 2025a), and DRIFTS (Zalevskyi et al., 2025b). We compare our Synthseg baseline with/without data-driven sampling (experiments 1 and 2) with the above papers. For a fairer comparison, we provide additional results that remove the effects of model emsembling: for each experiment, all runs are individually evaluated, and the results are averaged. The numbers are shown in table 3. In supplementary tables 6 and 7, we provide more detailed results for this non-ensemble version, with all experiments broken down by domain and by all underlying datasets that form each domain.

Our experiments are most comparable with FetalSynthSeg (its “KISPI-all” model), as they are trained on the same dataset and both methods employ fully synthetic training (i.e., no real MRIs are used as training input). FetalSynthSeg achieved significantly higher results than our experiments in the in-domain subjects, demonstrating a strong improvement to the synthseg baseline. On the similar-domain and out-of-domain, FetalSynthSeg performed on par with experiment 1, outperforming our data-driven sampling without ensembling by 0.03 and 0.02, respectively. On the most-pathological domain, FetalSynthSeg beats the non-ensemble experiment 1 by 0.03 but is outperformed by our data-driven sampling by 0.04 without ensembling and 0.07 with ensembling. We see that although FetalSynthSeg demonstrated its clear improvements compared to Synthseg in both in-domain subjects and subjects with severe shape variations, since its augmentations do not explicitly take shape variations into consideration, our data-driven sampler is capable of surpassing FetalSynthSeg in the most pathological cases despite only trained on the plain Synthseg pipeline.

Table 3: Comparing with more related works.
Dice scores of all testing subjects in each domain, all structures averaged for each subject (mean ± standard deviation). We provide additional non-ensemble results (referred to as “average”): for each experiment, each run is individually evaluated with the final metric averaged across all runs.

| Experiment | in-domain | similar-domain | out-of-domain | most-pathological |
|---|---|---|---|---|
| Exp 1: baseline (average) | 0.73±0.09 | 0.79±0.11 | 0.74±0.12 | 0.57±0.22 |
| Exp 1: baseline (ensemble) | 0.74±0.09 | 0.80±0.11 | 0.76±0.11 | 0.60±0.21 |
| Exp 2: samp (average) | 0.71±0.09 | 0.76±0.11 | 0.73±0.09 | 0.64±0.16 |
| Exp 2: samp (ensemble) | 0.73±0.09 | 0.77±0.11 | 0.75±0.08 | 0.67±0.15 |
| FetalSynthSeg (2024) | 0.80±0.09 | 0.79±0.10 | 0.75±0.13 | 0.60±0.22 |
| FeTA 2021 best † | 0.79±0.16 | n/a | n/a | n/a |
| FeTA 2022 best * † | 0.77±0.18 | n/a | n/a | n/a |
| FeTA 2024 best * † | 0.78±0.15 | n/a | n/a | n/a |
| DRIFTS * † (2025) | 0.78±0.09 | 0.85±0.09 | 0.84±0.05 | 0.81±0.08 |

* Trained on additional subjects with severe ventriculomegaly that overlaps with our out-of-domain and most-pathological test domains.
† Real MR Images are used for training (i.e., not a fully synthetic training pipeline).

We also provide the in-domain numbers for the best results of FeTA 2021, 2022, and 2024, though it should be noted that these pipelines incorporate real MR images for training. This is different from FetalSynthSeg and our method, which only uses the label maps. Similar to FetalSynthSeg, those results outperform our experiments on the in-domain test set. However, it is important to note that our goal isn’t to achieve the best results in a certain scenario, but to achieve strong generalizability under various image variations in shape, quality, modality, and appearances.

Last, we run DRIFTS, a very recent advancement, on our test domains as a reference. DRIFTS integrated synthetic training with real-images-based fine-tuning and weight interpolations, achieving an all-round improvement in a wide range of test sets compared to existing works. We see that DRIFTS outperforms our experiments in all test domains. However, it should be noted that its publicly available model is trained on additional subjects that directly overlaps with our out-of-domain and most-pathological test domains. Although trained on different datasets, the clear advantages of DRIFTS point the way to future work in fitting data-driven sampling to such training pipelines that utilize real images, where sampling could address not just shape-based variations but potentially intensity-related variations. This is especially possible since DRIFTS does not attempt to address domain shift through train-time sampling, which makes those two methods inherently complementary rather conflicting.

### 4.4 Clinical implications

Since its clinical use in the late 90s (Garel et al., 2001), fetal MRI has provided important insights into human fetal brain development. Besides its implementation as a clinical problem-solving tool in adjunct to fetal neurosonography (Prayer et al., 2023), it offered excellent tissue resolution of diverse fetal brain compartments, opening new research paths in quantitative fetal neuroimaging. Due to the expertise of the team performing fetal MRI examinations and its relatively limited availability at a high level of proficiency, the amount of fetal MRI data is still relatively small (Di Mascio et al., 2022) compared to other fields of developmental neuroimaging. Moreover, the fetal brain is a highly dynamic structure, changing its appearance on a daily to weekly basis. In addition, pathological alterations – such as ventriculomegaly – may alter the overall appearance of the fetal brain in a way that segmentation algorithms trained on normative data would produce inconclusive and useless results. However, all fetal quantitative neuroimaging approaches require a certain level of robustness to offer clinically valid reference data, which can be interpreted in the setting of neurodevelopmental outcome research. All these aspects strongly emphasize the need for other strategies to optimize fetal brain segmentation approaches, rather than simply relying on limited amounts of existing imaging data.

## 5 Conclusion and future work

For clinical use, fetal MR image segmentation algorithms must be reproducible, robust to image contrast variations, and effective across different MRI field strengths and sequences. These algorithms must also handle pathological cases that deviate from anatomical structures in training data. In this work, we introduced a novel data-driven strategy for train-time template sampling with the aim of increasing domain generalizability of trained networks. Our method is applied to SynthSeg and tested with other template augmentation methods on a diverse range of training/testing datasets. Our experiments showed notable improvements in cases with severe anatomical variations, though with a slighter performance decrease in cases with more typical anatomical structures. While SynthSeg used in fetal MR images demonstrated reliable performance across domain shifts in acquisition methods, sequences, field strengths, and reconstruction techniques, further work is needed to represent a broader range of pathological structural changes beyond ventriculomegaly. Future work also lies in combining our sampling strategy with other training pipelines for MR image segmentation networks, such as nnu-net, FetalSynthSeg, and DRIFTS.

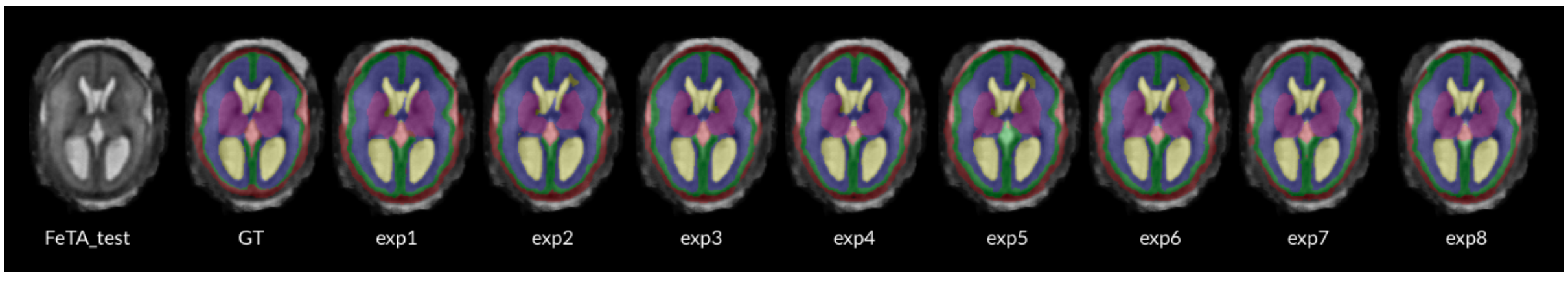

Figure 9: Segmentation visualization -- FeTA test
An example of the segmentation performance of each experiment on a testing subject from the in-domain FeTA 2021 testing set

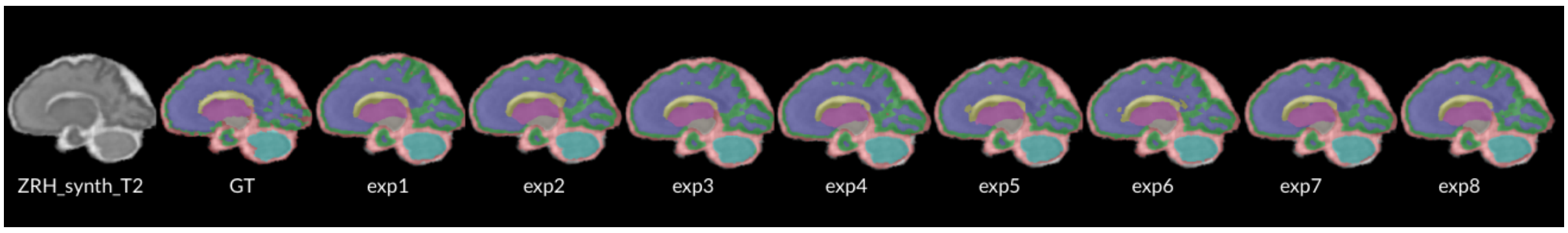

Figure 10: Segmentation visualization -- ZRH synth T2
An example of the segmentation performance of each experiment on a testing subject from the similar-domain ZRH synth T2 set.

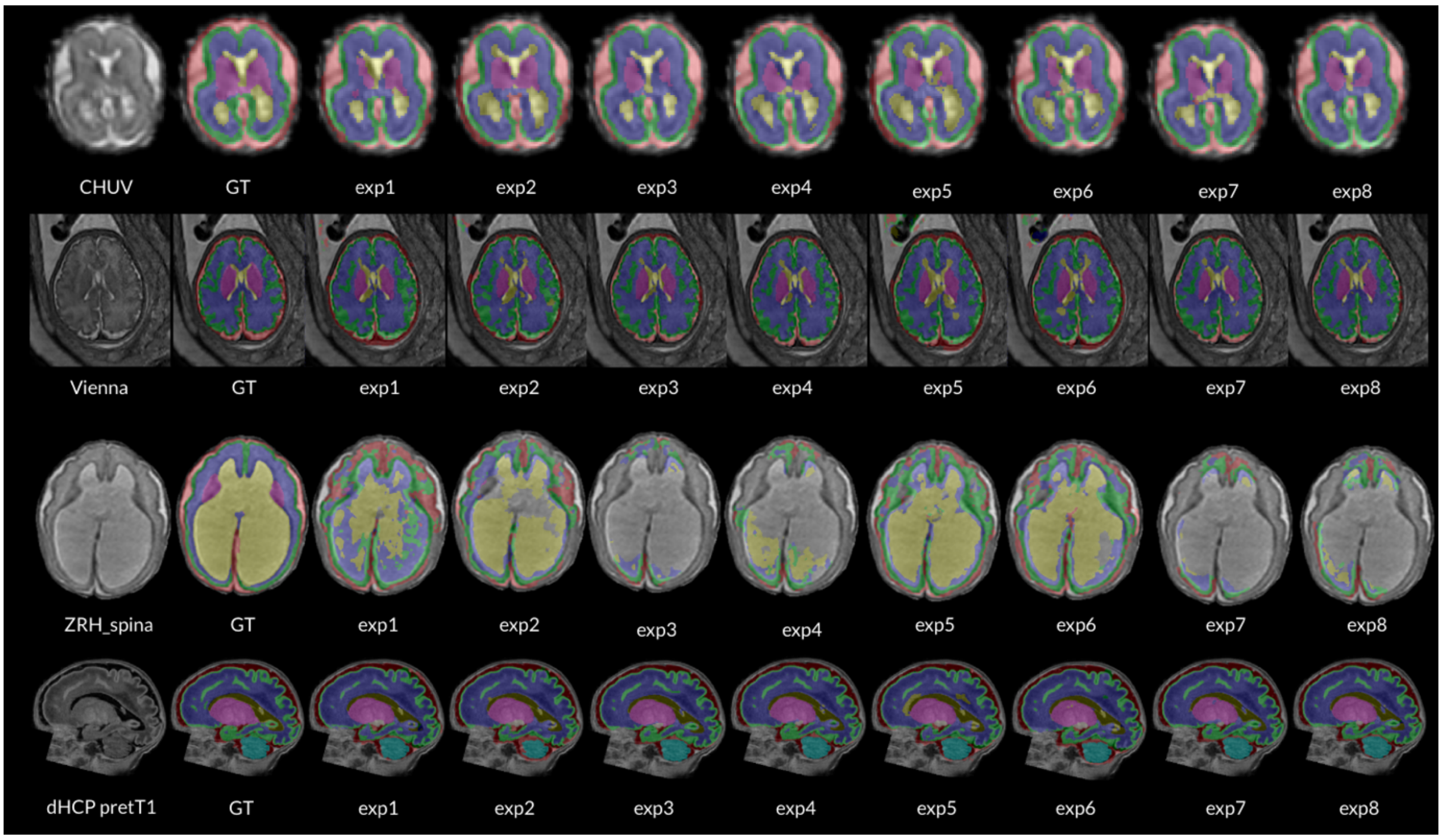

Figure 11: Segmentation visualization -- out-of-domain
Examples of the segmentation performance of each experiment on four testing subjects from the out-of-domain testing sets. The four subjects come from, respectively: CHUV, Vienna, ZRH spina (pathological), and dHCP_predT1.

## Highlights

- We address domain shift in brain segmentation models trained by Synthseg.
- We created a sampling strategy to alleviate class imbalance in training data.
- The strategy improved outcomes on subjects with intense anatomical abnormalities.
- The strategy is combined with many existing image augmentation schemes.
- All methods are thoroughly tested on diverse sets of training/testing data.

### Data and Code Availability Statement

This paper uses both open and private datasets from various sources. Relevant details are provided in Section 2.1.1, where open datasets are cited and could be requested by contacting the respective data providers. The code for this paper is publicly available at https://github.com/ZiyaoShang/synthseg-for-clinical-fetal-brains.

## Declaration of Generative AI and AI-assisted Technologies in the writing process

During the preparation of this work the authors used ChatGPT in order to enhance the language of this article. After using this tool, the authors reviewed and edited the content as needed and take full responsibility for the content of the published article.

## Studies in humans

All procedures were performed in compliance with relevant laws and institutional guidelines and have been approved by the appropriate institutional committees. The privacy rights of human subjects have been observed and that informed consent was obtained for experimentation with human subjects. Details of the ethical approvals are provided in Section 2.7.

## Acknowledgements and Funding Sources

This project was supported by the Swiss National Science Foundation, grant Nr. IZKSZ3_218590 and 10003124, the Adaptive Brain Circuits in Development and Learning Project, University Research Priority Program of the University of Zürich; by the Vontobel Foundation; by the Anna Müller Grocholski Foundation and the Prof. Max Cloetta Foundation.

The dHCP data were provided by the developing Human Connectome Project, KCL-Imperial-Oxford Consortium funded by the European Research Council under the European Union Seventh Framework Programme (FP/2007-2013) / ERC Grant Agreement no. [319456]. We are grateful to the families who generously supported this trial.

We acknowledge access to the facilities and expertise of the CIBM Center for Biomedical Imaging, a Swiss research center of excellence founded and supported by CHUV, UNIL, EPFL, UNIGE and HUG. This work is supported by Era-net Neuron MULTIFACT — Swiss National Science Foundation (SNSF) grant 31NE30_203977 and SNSF grants 182602 and 215641.

The low field KCL data was possible through funding from the UKRI [MR/T018119/1], DFG [502024488] and ERC [101165242].

NDA study DOI: 10.15154/5tzp-xm67 .

## Supplementary Material

| Experiment | CSF | GM | WM | LV | CB | TH | BS |
|---|---|---|---|---|---|---|---|
| Experiment 1: baseline | 0.67 | 0.63 | 0.86 | 0.78 | 0.80 | 0.71 | 0.73 |
| Experiment 2: samp | 0.65 | 0.61 | 0.85 | 0.77 | 0.79 | 0.70 | 0.73 |
| Experiment 3: dhcp | 0.67 | 0.64 | 0.84 | 0.77 | 0.78 | 0.71 | 0.72 |
| Experiment 4: dhcp+samp | 0.66 | 0.62 | 0.83 | 0.77 | 0.76 | 0.68 | 0.69 |
| Experiment 5: synth | 0.62 | 0.60 | 0.85 | 0.76 | 0.79 | 0.70 | 0.73 |
| Experiment 6: synth+samp | 0.65 | 0.60 | 0.84 | 0.77 | 0.79 | 0.70 | 0.72 |
| Experiment 7: dhcp+synth | 0.66 | 0.63 | 0.84 | 0.78 | 0.79 | 0.68 | 0.67 |
| Experiment 8: dhcp+synth+samp | 0.65 | 0.63 | 0.84 | 0.77 | 0.77 | 0.68 | 0.67 |

Supplementary Table 1: Dice scores: in-domain

The average Dice score for each structure among all subjects in the in-domain test set

| Experiment | CSF | GM | WM | LV | CB | TH | BS |
|---|---|---|---|---|---|---|---|
| Experiment 1: baseline | 0.77 | 0.76 | 0.88 | 0.72 | 0.88 | 0.80 | 0.80 |
| Experiment 2: samp | 0.72 | 0.72 | 0.86 | 0.64 | 0.87 | 0.80 | 0.80 |
| Experiment 3: dhcp | 0.78 | 0.78 | 0.89 | 0.75 | 0.88 | 0.82 | 0.82 |
| Experiment 4: dhcp+samp | 0.78 | 0.77 | 0.88 | 0.71 | 0.88 | 0.81 | 0.84 |
| Experiment 5: synth | 0.74 | 0.72 | 0.85 | 0.59 | 0.88 | 0.79 | 0.79 |
| Experiment 6: synth+samp | 0.77 | 0.73 | 0.87 | 0.63 | 0.88 | 0.80 | 0.80 |
| Experiment 7: dhcp+synth | 0.78 | 0.77 | 0.88 | 0.71 | 0.88 | 0.82 | 0.86 |
| Experiment 8: dhcp+synth+samp | 0.77 | 0.77 | 0.88 | 0.71 | 0.88 | 0.82 | 0.85 |

Supplementary Table 2: Dice scores: similar-domain

The average Dice score for each structure among all subjects in the similar-domain test set

| Experiment | CSF | GM | WM | LV | CB | TH | BS |
|---|---|---|---|---|---|---|---|
| Experiment 1: baseline | 0.70 | 0.69 | 0.84 | 0.75 | 0.86 | 0.73 | 0.77 |
| Experiment 2: samp | 0.67 | 0.67 | 0.84 | 0.71 | 0.84 | 0.76 | 0.79 |
| Experiment 3: dhcp | 0.70 | 0.71 | 0.83 | 0.75 | 0.83 | 0.73 | 0.78 |
| Experiment 4: dhcp+samp | 0.70 | 0.70 | 0.81 | 0.74 | 0.82 | 0.71 | 0.79 |
| Experiment 5: synth | 0.66 | 0.66 | 0.83 | 0.66 | 0.85 | 0.73 | 0.77 |
| Experiment 6: synth+samp | 0.68 | 0.67 | 0.82 | 0.67 | 0.86 | 0.74 | 0.79 |
| Experiment 7: dhcp+synth | 0.68 | 0.71 | 0.82 | 0.72 | 0.85 | 0.73 | 0.82 |
| Experiment 8: dhcp+synth+samp | 0.68 | 0.71 | 0.83 | 0.72 | 0.83 | 0.72 | 0.79 |

Supplementary Table 3: Dice scores: out-of-domain

The average Dice score for each structure among all subjects in the out-of-domain test set

| Experiment | CSF | GM | WM | LV | CB | TH | BS |
|---|---|---|---|---|---|---|---|
| Experiment 1: baseline | 0.56 | 0.50 | 0.62 | 0.77 | 0.71 | 0.47 | 0.61 |
| Experiment 2: samp | 0.55 | 0.52 | 0.69 | 0.88 | 0.75 | 0.57 | 0.73 |
| Experiment 3: dhcp | 0.46 | 0.46 | 0.54 | 0.64 | 0.54 | 0.46 | 0.54 |
| Experiment 4: dhcp+samp | 0.48 | 0.45 | 0.55 | 0.77 | 0.56 | 0.42 | 0.58 |
| Experiment 5: synth | 0.53 | 0.50 | 0.69 | 0.90 | 0.70 | 0.49 | 0.64 |
| Experiment 6: synth+samp | 0.57 | 0.51 | 0.68 | 0.88 | 0.75 | 0.54 | 0.69 |
| Experiment 7: dhcp+synth | 0.46 | 0.44 | 0.55 | 0.74 | 0.67 | 0.44 | 0.61 |
| Experiment 8: dhcp+synth+samp | 0.47 | 0.47 | 0.56 | 0.71 | 0.57 | 0.39 | 0.54 |

Supplementary Table 4: Dice scores: most-pathological

The average Dice score for each structure among all subjects in the most-pathological test set.

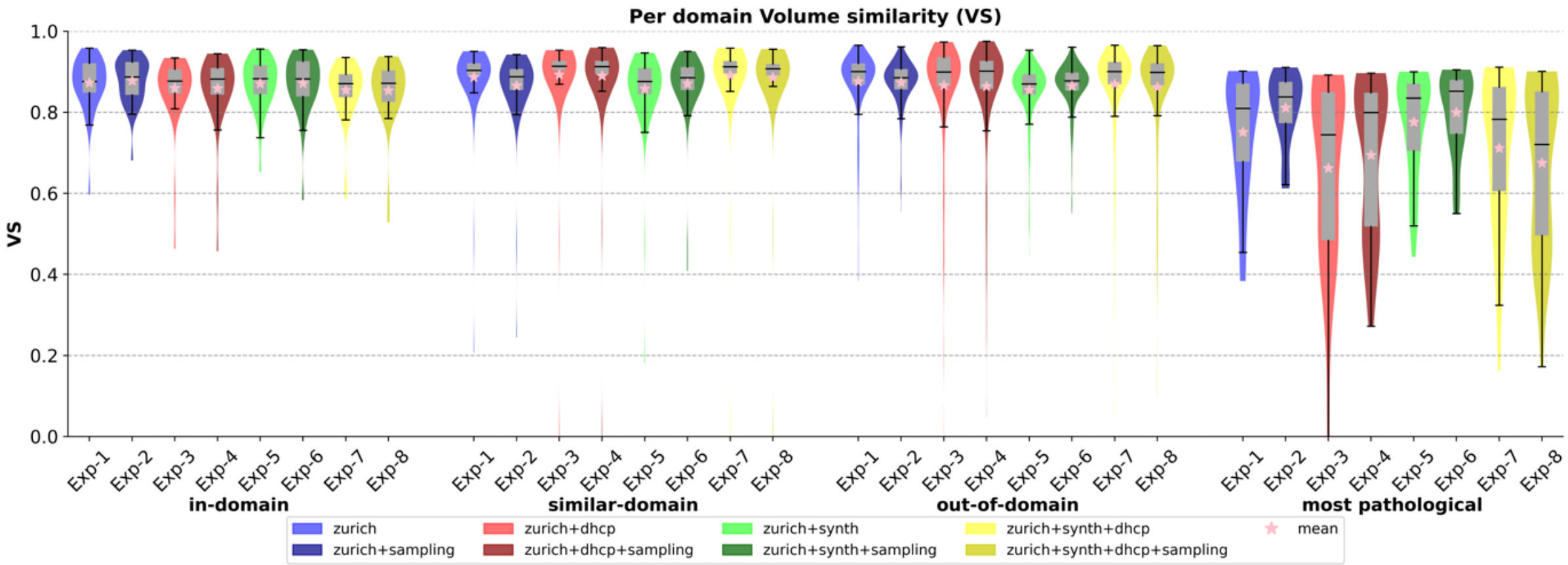

Supplementary Figure 1: Comparison of Volume Similarity (VS) between domains.

For each domain, we plot the distribution of the VS mean of all structures for every subject.

| Test set | Exp1 | Exp2 | Exp3 | Exp4 | Exp5 | Exp6 | Exp7 | Exp8 |
|---|---|---|---|---|---|---|---|---|
| FeTA_test | 0.74 | 0.73 | 0.73 | 0.72 | 0.72 | 0.73 | 0.72 | 0.71 |
| ZRH_new_crtl | 0.80 | 0.77 | 0.82 | 0.81 | 0.74 | 0.77 | 0.82 | 0.81 |
| ZRH_synth_T2 | 0.80 | 0.78 | 0.81 | 0.81 | 0.80 | 0.80 | 0.81 | 0.82 |
| dHCP_pretT1 | 0.78 | 0.75 | 0.81 | 0.80 | 0.74 | 0.75 | 0.80 | 0.80 |
| CHUV | 0.79 | 0.78 | 0.81 | 0.80 | 0.76 | 0.77 | 0.80 | 0.80 |
| ZRH_synth_T1 | 0.78 | 0.75 | 0.78 | 0.78 | 0.78 | 0.77 | 0.78 | 0.78 |
| dHCP_pretT2 | 0.80 | 0.76 | 0.83 | 0.80 | 0.75 | 0.76 | 0.81 | 0.80 |
| KCL | 0.75 | 0.74 | 0.77 | 0.72 | 0.71 | 0.73 | 0.75 | 0.76 |
| Vienna | 0.60 | 0.65 | 0.58 | 0.61 | 0.57 | 0.61 | 0.61 | 0.58 |
| ZRH_Spina | 0.79 | 0.79 | 0.76 | 0.75 | 0.79 | 0.79 | 0.77 | 0.76 |
| Most_pathological | 0.60 | 0.67 | 0.52 | 0.55 | 0.64 | 0.66 | 0.56 | 0.53 |

Supplementary Table 5: Dice scores: all base datasets.

We separate all underlying datasets of which each test domain is composed and report their results separately for every experiment.

| Experiment (no ensemble) | in-domain | similar-domain | out-of-domain | most-pathological |
|---|---|---|---|---|
| Experiment 1: baseline | **0.73**±0.09 | 0.79±0.11 | 0.74±0.12 | 0.57±0.22 |
| Experiment 2: samp | 0.71±0.09 | 0.76±0.11 | 0.73±0.09 | **0.64**±0.16 |
| Experiment 3: dhcp | 0.72±0.11 | **0.81**±0.12 | **0.75**±0.14 | 0.52±0.24 |
| Experiment 4: dhcp+samp | 0.70±0.12 | 0.80±0.12 | 0.74±0.13 | 0.54±0.21 |
| Experiment 5: synth | 0.71±0.08 | 0.76±0.11 | 0.71±0.11 | 0.61±0.20 |
| Experiment 6: synth+samp | 0.71±0.09 | 0.77±0.10 | 0.72±0.10 | 0.62±0.18 |
| Experiment 7: dhcp+synth | 0.70±0.10 | 0.80±0.11 | 0.74±0.13 | 0.55±0.21 |
| Experiment 8: dhcp+synth+samp | 0.70±0.11 | 0.80±0.12 | 0.74±0.15 | 0.51±0.24 |

Supplementary Table 6: Dice scores result summary (no ensemble).

Dice scores among all testing subjects in each domain, all structures averaged for each subject (mean ± standard deviation). No model emsembling is performed in this table, the matrices are the average of the three runs for each experiment.

| Test set | Exp1 | Exp2 | Exp3 | Exp4 | Exp5 | Exp6 | Exp7 | Exp8 |
|---|---|---|---|---|---|---|---|---|
| FeTA_test | 0.73 | 0.71 | 0.72 | 0.70 | 0.71 | 0.71 | 0.70 | 0.70 |
| ZRH_new_crtl | 0.78 | 0.75 | 0.81 | 0.79 | 0.73 | 0.75 | 0.80 | 0.79 |
| ZRH_synth_T2 | 0.80 | 0.77 | 0.81 | 0.80 | 0.80 | 0.80 | 0.81 | 0.81 |
| dHCP_pretT1 | 0.76 | 0.72 | 0.80 | 0.78 | 0.72 | 0.73 | 0.79 | 0.79 |
| CHUV | 0.78 | 0.76 | 0.80 | 0.79 | 0.74 | 0.75 | 0.79 | 0.79 |
| ZRH_synth_T1 | 0.77 | 0.73 | 0.77 | 0.76 | 0.77 | 0.76 | 0.76 | 0.77 |
| dHCP_pretT2 | 0.78 | 0.74 | 0.81 | 0.79 | 0.74 | 0.75 | 0.79 | 0.79 |
| KCL | 0.72 | 0.70 | 0.75 | 0.70 | 0.67 | 0.68 | 0.73 | 0.74 |
| Vienna | 0.55 | 0.60 | 0.57 | 0.61 | 0.51 | 0.54 | 0.59 | 0.55 |
| ZRH_Spina | 0.77 | 0.76 | 0.75 | 0.74 | 0.77 | 0.77 | 0.75 | 0.74 |
| Most_pathological | 0.57 | 0.64 | 0.52 | 0.54 | 0.61 | 0.62 | 0.55 | 0.51 |

Supplementary Table 7: Dice scores--all base datasets (no emsemble).

We separate all underlying datasets of which each test domain is composed and report their results separately for every experiment. No model emsembling is performed in this table, the matrices are the average of the three runs for each experiment.

## Supplementary Note 1: Effective percentage of regular templates

For the analysis in the last paragraph of Section 3.4. We used the ratio of regular/pathological templates present in the training set for each experiment. A ratio calculated by simply adding up the numbers of regular and pathological templates is not the most accurate estimation because, as mentioned in Section 2.6, we add an additional constraint such that the sum of the sampling weights of all templates in dHCP_fetal always adds up to 50%. As a result, the accurate percentage of regular templates (i.e., “effective percentage”) is a weighted sum of the percentages for [dHCP_fetal] and [FeTA_train + Synthetic templates]. This gives:

**Experiments 1 and 2:**
31/80 = 39%

**Experiments 3 and 4:**
[31/80] * 0.5 + 1 * 0.5 = 69%

**Experiments 5 and 6:**

31/(80+80) = 19%

**Experiments 7 and 8:**
[31/(80+80)] * 0.5 + 1 * 0.5 = 60%